\documentclass[aps,prb,twocolumn,superscriptaddress]{revtex4-2}
\usepackage[latin1]{inputenc}
\usepackage{graphicx}
\usepackage{float}
\usepackage{amsmath,amsfonts}
\usepackage{xcolor}
\usepackage{bbold}
\usepackage[colorlinks=true,allcolors=blue]{hyperref}
\usepackage{ulem}

\begin{document}
\title{Microscopic analysis of spin-momentum locking on a geometric phase metasurface}

\author{Fernando Lor\'en}
\affiliation{Instituto de Nanociencia y Materiales de Arag\'on (INMA), CSIC-Universidad de Zaragoza, 50009 Zaragoza, Spain\looseness=-1}
\affiliation{Departamento de F\'isica de la Materia Condensada, Universidad de Zaragoza, 50009 Zaragoza, Spain\looseness=-1}

\author{Gian L. Paravicini-Bagliani}
\affiliation{University of Strasbourg and CNRS, CESQ \& ISIS (UMR 7006), 8, all\'ee G. Monge, 67000 Strasbourg, France}

\author{Sudipta Saha}
\affiliation{University of Strasbourg and CNRS, CESQ \& ISIS (UMR 7006), 8, all\'ee G. Monge, 67000 Strasbourg, France}

\author{J\'er\^ome Gautier}
\affiliation{University of Strasbourg and CNRS, CESQ \& ISIS (UMR 7006), 8, all\'ee G. Monge, 67000 Strasbourg, France}

\author{Minghao Li}
\affiliation{University of Strasbourg and CNRS, CESQ \& ISIS (UMR 7006), 8, all\'ee G. Monge, 67000 Strasbourg, France}

\author{Cyriaque Genet} 
\email{genet@unistra.fr}
\affiliation{University of Strasbourg and CNRS, CESQ \& ISIS (UMR 7006), 8, all\'ee G. Monge, 67000 Strasbourg, France}

\author{Luis Mart\'in-Moreno}
\email{lmm@unizar.es} 
\affiliation{Instituto de Nanociencia y Materiales de Arag\'on (INMA), CSIC-Universidad de Zaragoza, 50009 Zaragoza, Spain\looseness=-1}
\affiliation{Departamento de F\'isica de la Materia Condensada, Universidad de Zaragoza, 50009 Zaragoza, Spain\looseness=-1}

\begin{abstract}
We revisit spin-orbit coupling in a plasmonic Berry metasurface comprised of rotated nanoapertures, which is known to imprint a robust far-field polarization response. We present a scattering formalism that shows how that spin-momentum locking emerges from the geometry of the unit cell without requiring global rotation symmetries. We find and confirm with Mueller polarimetry measurements that spin-momentum locking is an approximate symmetry.  The symmetry breakdown is ascribed to the elliptical projection of circularly polarized light into the planar surface. This breakdown is maximal when surface waves are excited, and a new set of spin-momentum locking rules is presented for this case.
\end{abstract}

\maketitle

\section{Introduction}
Chiral light-matter interactions~\cite{bliokh2015quantum, lodahl2017chiral} form the core of recent discussions in quantum optics and material science. 
Recently,  such interactions have been investigated in nano-optics, with the appropriate designs of two-dimensional nanoantennas or plasmonic arrays~\cite{shitrit2013spinoptical,fox2022generalized,bliokh2015spinorbit}. Chiral metasurfaces, sometimes known as ``Berry'' or ``geometric phase metasurfaces" (GPM)~\cite{bomzon2002spacevariant}, have found many applications for the selective manipulation of quantum emitters, in particular in the field of valleytronics~\cite{sun2019separation, guddala2019valley, jha2018spontaneous}. There, valley excitons can be selectively excited and detected by the spin angular momentum of the emitted light due to the metasurface's spin-momentum locking (SML) mechanism. This allows routing valley degrees of freedom into optical cavity modes, opening ways for new valley-photon interfaces~\cite{gong2018nanoscale, chervy2018room, rong2020photonic}. 

Despite their applicative potential and the fascinating connections they draw with many fundamental issues in optics, plasmonic GPMs have been elusive to a rigorous and exhaustive theoretical description. 
Previous theoretical works have either considered (i) Berry-phase arguments in systems with optical elements presenting a {\it continuous} spatial modulation~\cite{bomzon2002spacevariant} or (ii) a group theory analysis in the Kagome lattice, restricted to waves with an electric field perpendicular to the surface, that ascribed SML to the simultaneous presence of translation and rotation symmetries of the {\it whole} lattice~\cite{shitrit2013spinoptical}.  
None of these approaches cover the typical case of GPMs, which are composed of discrete elements that present chirality within the unit cell but without global rotation symmetries~\cite{genevet2017recent, chervy2018room, fox2022generalized}. 

Another issue that has not yet been addressed is how SML, which distinguishes between two circular polarization states, can be reconciled with the surface plasmon polariton (SPP), which only exists for TM polarization. In other words, a strict SML rule implies that all Bragg modes, including surface waves, must have a well-defined circular polarization, which is not possible when the Bragg mode is an SPP.

In this article, we propose a scattering formulation that clarifies how SML appears in a plasmonic holey GPM. Our formulation has three strong assets with respect to previous theoretical works. First, it shows that the SML appears from the geometry of the unit cell. Second, it demonstrates that the SML rules are only approximate when predicting polarization, due to the elliptical projection onto the GPM of the circularly polarized diffraction orders.  Third, it enables a direct comparison with experimental data.  The validity of our findings is confirmed by the excellent agreement between theory and new experimental data of the polarization states associated with each of the diffracted Bragg modes, determined via Mueller polarimetry performed in the optical Fourier space.

\begin{figure}[h]
    \includegraphics[width=\columnwidth]{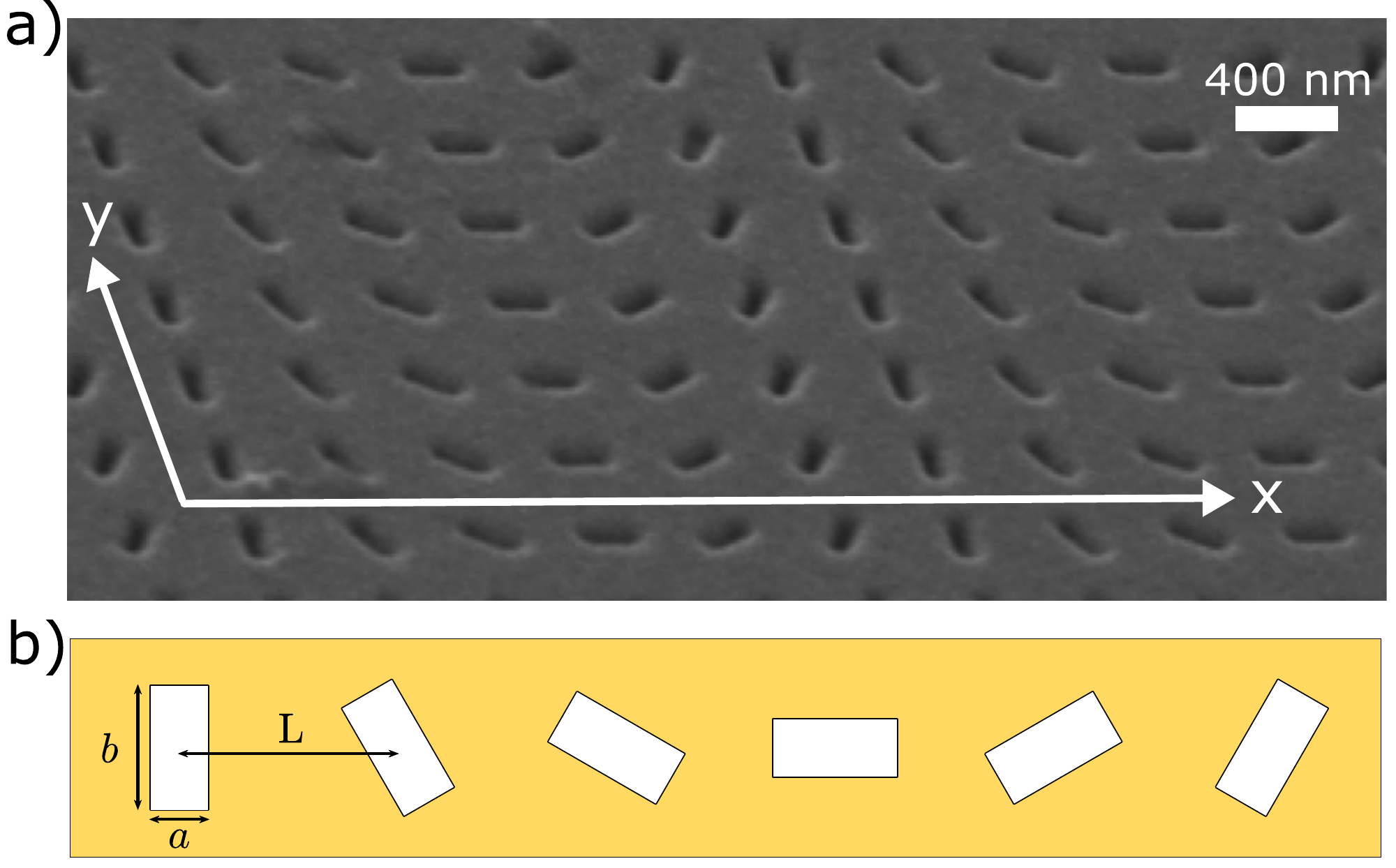}
    \caption{(a) Scanning electron microscope (SEM) image of the periodic metasurface, composed of rectangular grooves in a gold film. (b) Schematic bird's-eye view of the unit cell. Nominal parameters in panel (a): $a = 80 $~nm, $b = 220 $~nm, $d = 60 $~nm, and $L = 460$~nm.}
        \label{fig:sample}
\end{figure} 

\section{Theoretical formalism}
We consider the simplest GPM, characterized by a periodic array of rectangular grooves in a metal slab. Figure~\ref{fig:sample}(a) presents an SEM image of the periodic gold metasurface that we will experimentally study later, and Fig.~\ref{fig:sample}(b) is a schematic representation of the unit cell, which comprises $N$ grooves, and is periodically repeated in both $\vec{u}_{x}$ and $\vec{u}_{y}$ directions. Each groove has a short side $a$, a long side $b$, and depth $d$.  The $\alpha$ th groove is rotated at an angle $\theta_{\alpha}$ with respect to the $\vec{u}_{x}$ axis.  

In this paper, we present a general theoretical framework for arbitrary groove positions and $\theta_{\alpha}$, but we will analyze the case where (i) the groove centers are aligned along the $\vec{u}_{x}$ axis, (ii) the distance between the centers of nearest grooves is $L$, in both $x$- and $y$- directions and (iii) $\theta_{\alpha}$ varies linearly with $\alpha$, $\theta_{\alpha} =  2 \pi n_w \alpha/N$. 

The winding number $n_w$ defines the number of complete $2\pi$ rotations along the unit cell (in the particular case $n_w = 0$, the array becomes a nonchiral square lattice of rectangular grooves). The system depicted in Fig.~\ref{fig:sample}(b) has $n_w= 1/2$, as each rectangular groove presents a mirror symmetry with respect to the middle of the long axis. Consequently, the rotation along one unit cell is only of $\pi$ radians because the next unit cell is the same and completes the $2\pi$ rotation. Notice that, generally, a surface characterized by a constant increment in angle $\theta$ from hole to hole requires an integer $n_w$. However, a rectangular aperture has reflection symmetry along its long axis, which makes the collection of holes repeat itself already for $n_w=1/2$, thus halving the unit cell size (if holes depicted in Fig.~\ref{fig:sample}(b) were trapezoidal, the unit cell would comprise 12 holes while the structure repeats itself already after 6 rectangular holes). Notice also that although the grooves perform a stepwise rotation within the unit cell, the whole lattice does not support a global rotation symmetry.  

We consider an electromagnetic plane wave impinging onto the structure with an in-plane wave vector along the $x$ direction, $\vec{k}^{in} = k_{x}^{in} \vec{u}_{x}$, and compute the reflection coefficients into the different Bragg modes that can be diffracted off the periodic metasurface. For that, we use the coupled-mode method (CMM), which has been widely used in the study of electromagnetic (EM) properties in holey metallic films~\cite{garciavidal2010light}.  The CMM expands the EM fields in plane waves in the free space regions and waveguide modes inside the grooves and finds the field amplitudes by adequately matching the fields at the interfaces. 
 
The equations derived from the CMM are usually written in terms of the amplitudes of the waveguide modes~\cite{MartinMoreno2001,lmm2008minimal,deleon2008theory}.  However, when studying the SML mechanism, we find it more convenient to derive the equations directly for the reflection coefficients. Although computationally less efficient, this provides a more transparent description as it now involves far-field amplitudes, thus avoiding the near-field to far-field decoding needed to extract scattering coefficients from waveguide mode amplitudes. 

Bragg modes are characterized by an in-plane momentum $\vec{k}= \vec{k}^{in} +  \vec{G}$, where $\vec{G}$ is a reciprocal lattice vector, in this case $\vec{G} = 2\pi m / (N L) \, \vec{u}_{x} +  2\pi n_{y} / L \, \vec{u}_{y}$, where $m$ and $n_y$ are integers. To develop a minimal model for studying the SML and simplify the presentation, in the main text, we consider only Bragg modes with $n_{y}=0$.  As shown in the Supp. Mat.~\cite{suppmat} Sec.~2,  this does not change the main results because along the $y$ axis there is no breaking of inversion symmetry.  We collect the two circular polarizations for the reflected $m$ th Bragg mode in the spinor $ \mathbf{r}_{m} \equiv (r_{m}^{+}, r_{m}^{-})^T$, where $\pm$ denote the right- and left-handed polarizations, each of them defined within the plane perpendicular to the wave vector of the corresponding Bragg mode, $\vec{k}_m$.  We choose the spin representation because the spin of a plane wave is conserved upon reflection by a mirror \cite{bliokh2013dual,cameron2012optical,cameron2014optical,cameron2017chirality} (while the helicity changes sign). 

The CMM can take into account the dielectric constant in the metal via the implementation of the surface impedance boundary conditions (SIBC).  The general expressions can be found in the Supp. Mat.~\cite{suppmat} Sec.~1, and are the ones used below when comparing to the experimental data. Here, we present the expressions for the simpler case in which the metal is treated as a perfect electric conductor (PEC), as this is sufficient to discuss the physics and the structure of the equations: 
\begin{equation}
    \begin{aligned}
  \mathbf{r}_{m} =  & - \delta_{m 0} \, \mathbf{i}_0 + C_{m 0} \,  Y_{0} \,  \mathbf{i}_0 - \sum_{m'} C_{m m'}  \, Y_{m'}  \, \mathbf{r}_{m'}.    \end{aligned}
    \label{eq:approxSML}
\end{equation}
The first term takes into account the specular reflection ($\mathbf{i}_0$ is the amplitude of the incident wave). The coefficients $C_{m m'}$, which we call ``geometric couplings'',  are $2\times2$ matrices operating in polarization space. They couple different Bragg modes via scattering with the GPM and encode the geometry of the unit cell through the overlaps between Bragg and waveguide modes (see the Supp. Mat.~\cite{suppmat} Sec.~1).  
This is, they contain the information on the geometric distribution of the holes in the unit cell and thus on whether they are rotated or not. $Y_{m}$ are the modal admittance matrices of the Bragg modes in the circular polarization basis (which relate the in-plane magnetic field to the electric one). They can be written as  
$Y_{m}= \bar{Y}_{m}\, \mathbb{1} +\Delta_{m} \,\sigma_{x}$, where $\mathbb{1}$ and $ \sigma_{x}$ are the $2\times2$ unit matrix and the Pauli matrix that swaps spin states, respectively.  In terms of the linear $p$ (transverse magnetic) - $s$ (transverse electric) polarized basis,  $\bar{Y}_{m} \equiv (Y_{mp}+Y_{ms})/2$ and $\Delta_{m} \equiv (Y_{mp}-Y_{ms})/2$.  For a plane wave with frequency $\omega$ and in-plane wave vector $k_{m}$ propagating in a uniform medium with dielectric constant $\epsilon$, $Y_{mp} = \epsilon/q_{mz}$ and $Y_{ms} = q_{mz}$, where $q_{mz} = \sqrt{\epsilon -  (c k_{m}/ \omega)^{2}}$ (c being the speed of light).  

Notice that $\Delta_{0}=0$ at normal incidence, while both $\bar{Y}_{m}$ and $\Delta_{m}$ diverge at the Rayleigh points (i.e., whenever a diffractive order becomes tangent to the metal-dielectric interface, as then $q_{mz} =0$).

\section{Spin-Momentum Locking}
The geometric couplings have a simple analytical expression when the polarization is defined on the circular basis with respect to the $\vec{u}_z$ direction ($C_{mm'}^z$). When the groove dimensions are much smaller than the wavelength, we find (see Supp. Mat.~\cite{suppmat} Sec.~1 and 3 for detailed derivation)
\begin{equation}
    C_{m m'}^{z} =   C\left( \delta_{m,m'+n_0 N} \, \mathbb{1} + \sum_{s=\pm} \delta_{m,m'+n_0 N-2 n_w s} \, \sigma_{s}  \right),
\label{eq:Zs}
\end{equation}
where $C$ is a constant that only depends on the properties of a single groove, $n_0$ is an integer, and $\sigma_{\pm}$ are Pauli matrices that increase and decrease spin, respectively.
The expression of the couplings coefficients in the circular polarization basis of each Bragg mode, $C_{mm'}$, can be obtained from $C_{mm'}^z$ using the $2 \times 2$ ``rotation" matrices $R^{k(m)\leftarrow z}$ and $R^{z\leftarrow k(m')}$:
\begin{equation}
  C_{m m'} =  R^{k(m)\leftarrow z} \, C_{m m'}^{z} \, R^{z\leftarrow k(m')}.
\end{equation}
The expressions for $R^{k(m)\leftarrow z}$  and $R^{z\leftarrow k(m')}$ are provided in the Supp. Mat.~\cite{suppmat} Sec.~1.

If all $\Delta_{m}$ were zero and all change of basis matrices $R$ were the identity, which only occurs at the direction normal to the surface, then the previous expressions would give rise to the following Bragg laws. The term proportional to $\mathbb{1}$ in the Eq.~\ref{eq:Zs} preserves spin, $\sigma_{out} = \sigma_{in}$, and the associated Bragg law $k_x^{out} = k_x^{in} +  n_0 G^{0} $, with $G^{0} = 2 \pi/L$, would be the same one that would appear if all grooves were parallel (and so it is denoted as ``standard'' Bragg law~\cite{shitrit2013spinoptical}). The terms inside the sum in the Eq.~\ref{eq:Zs} swap spin, $\sigma_{out} = \sigma_{in} \pm 1$, and shift the standard Bragg law by a term that depends both on the spin change and the winding number: $k_x^{out} = k_x^{in} + n_0 G^0 \mp k_{g}$, where $k_{g } \equiv 2\pi \, 2 n_w /(NL)$ is the geometric momentum. This condition is denoted as spin-orbit Bragg law~\cite{shitrit2013spinoptical} and corresponds to the exact SML mechanism.

These two Bragg laws have been amply used to discuss experimental results but, as mentioned before, they were derived only for the cases of continuous spatial modulation~\cite{bomzon2002spacevariant} and in a lattice that presents a combination of translation and rotation symmetry of the {\it whole} lattice)~\cite{shitrit2013spinoptical}. In our treatment, they appear from the groove-mediated geometric couplings between Bragg modes in a system without a global rotational lattice symmetry. Thus, SML is a feature of the basis of the unit cell and not of the symmetry of the whole lattice.

However, $\Delta_{m} \ne 0$ and $R \neq \mathbb{1}$ for Bragg modes with wave vectors away from the surface normal, so the terms proportional to $\sigma_{x}$ in both $Y_m$ and the modifications when passing from $C_{mm'}^z$ to $C_{mm'}$ must be considered. These terms flip circular polarization before the geometric couplings are applied. Thus the symmetry that leads to Eq.~\ref{eq:Zs} still holds,  but the exact link between changes in momentum and spin, found when assuming $\Delta_{m} = 0$ and $R=\mathbb{1}$, breaks down.

The physical origin of the breakdown terms resides in that the polarization of the transversal EM field is defined on the plane {\it perpendicular to the wave vector}. However, the in-plane component (i.e. \textit{perpendicular to the surface normal}) of the EM field is the relevant one in the interaction with the holey metasurface. This mismatch results in the EM wave being elliptically polarized and thus described by a combination of the two circular polarizations with respect to the propagation direction.
 
Mathematically,  for a PEC, the breakdown term is maximum for waves with $q_{mz}=0$, while for a real metal, this occurs when the Bragg mode coincides with the SPP \textit{of the flat surface}. Considering SML breakdown terms is thus essential when surface resonances are excited (spoof SPPs in the case of a PEC, SPPs of the corrugated surface in the case of a real metal \cite{pendry2004mimicking}). 
\begin{figure}
    \includegraphics[width=\columnwidth]{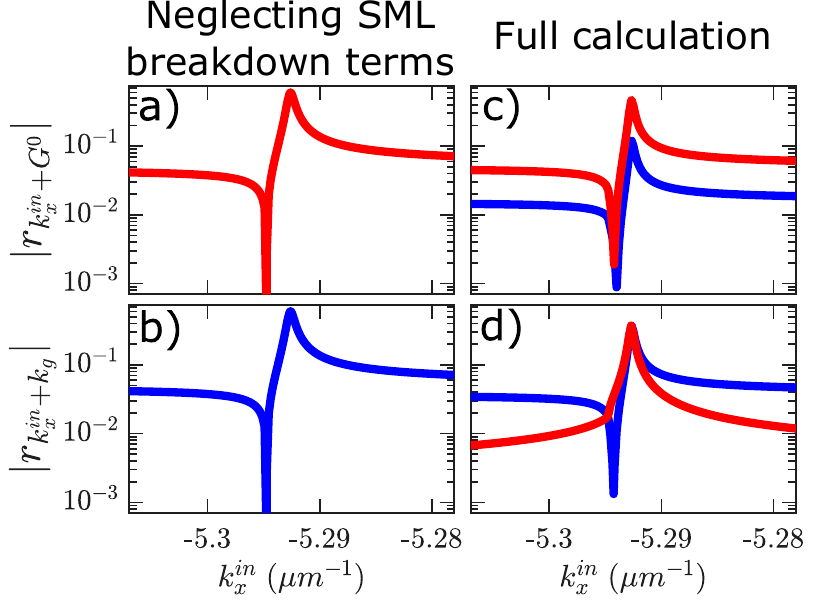}
    \caption{Reflection coefficients for the two polarizations  (spin $+$ in red and spin $-$ in blue) and two different diffraction orders.  The incident EM wave has spin $+$ and in-plane momentum $k_x^{in}$. The chosen resonance is such that $k_{x}^{SPP} \approx  k_x^{in} + G^0 +k_g$.  (a, b) Approximate calculation neglecting SML breakdown terms. $r_{ k_x^{in} + G^0}=0$ for spin $-$ in panel (a) and $r_{ k_x^{in} + k_g}=0$ for spin $+$ in panel (b). (c, d) Full calculation, including breakdown terms.  Chosen values: $\omega = 2.1 \; eV$, $a = 80 $~nm, $b = 220 $~nm, $d = 60 $~nm, $L = 460 $~nm, $n_w = 1/2$ and $N=6$, based on the experimental sample shown in Fig.~\ref{fig:sample}. The metal is considered as a PEC.}
        \label{fig:appSML}
\end{figure}

To illustrate the relevance of the breakdown terms, we consider an incident left-handed (spin $+$) polarized wave, impinging onto the GPM described in Fig.~\ref{fig:sample}. We compute the reflection coefficients at a fixed frequency as a function of $k_x^{in}$, near an SPP resonance. 
Figure~\ref{fig:appSML} renders the results for two different Bragg orders ($r_{k_x^{in} + G^0}$ and $r_{k_x^{in}+k_g}$) and the two circular polarizations (spin $+$ in red and spin $-$ in blue).  
The left panels (a and b) are computed neglecting all breakdown terms by artificially forcing both $\Delta_{m}=0$ (taking $Y_{ms} = Y_{mp}$) and  $C_{m m'} =  C_{m m'}^{z} $, while the right panels (c and d) are the full calculations including the breakdown terms.
As expected, when breakdown terms are neglected, the coefficient $r_{k_x^{in} + G^0}$ is nonzero only for spin $+$, while the Bragg mode that has gained an extra geometric momentum $k_{g}$ has an associated spin reduction, and is thus nonzero only for spin $-$. When the breakdown terms are considered, as they should, both polarizations become finite for any Bragg order. Away from the SPP resonance, the effect of the breakdown terms is small and SML selection rules hold to a good approximation. However, the breakdown terms can not be neglected at resonance. So, a new set of phenomenological selection rules is needed to understand the SML in plasmonic metasurfaces when SPPs are excited. This will be analyzed in detail in the next section when describing the excitation and deexcitation of SPPs.

Notice that we can apply the geometric couplings sequentially. This is, we can study the couplings between a Bragg mode and a second one, and then the couplings between the latter and another different.  This way, the first mode can be coupled to modes with two or more units of added geometric momentum ($\pm 2 \, k_g, \pm 3 \, k_g$, etc.). Of course, this would not be possible if the SML were perfect because spinor algebra would prevent this process (it is impossible to reduce more than one spin unit to a spin 1/2). However, as we will show in the next section, breakdown terms enable this remarkable phenomenology.


\section{Mueller polarimetry}
A more detailed analysis of the metasurface's full polarization response can be obtained with Mueller polarimetry that measures, artifact-free, the polarization states of the light beams incident on and scattered off the metasurface.  The $4 \times 4$ Mueller matrix $M$ is defined by $\mathbf{S}_{out}=M \, \mathbf{S}_{in}$, where $\mathbf{S}_{in/out}$ are the input and output polarization states described by the Stokes vectors~\cite{menzel2010advanced,gil2017polarized,fujiwara2007spectroscopic}. 
Using the experimental scheme presented in the Supp. Mat. \cite{suppmat} Sec.~4, $M$ can be measured as a function of photon energy $\omega$ and $k_x^{in}$. 
Among the 16 components of the Mueller matrix, we concentrate on the component $M_{30}$, which provides the difference in reflected intensities between the $+$ and $-$ polarizations when the system is illuminated with unpolarized light. 

\begin{figure}    
	\includegraphics[width=\columnwidth]{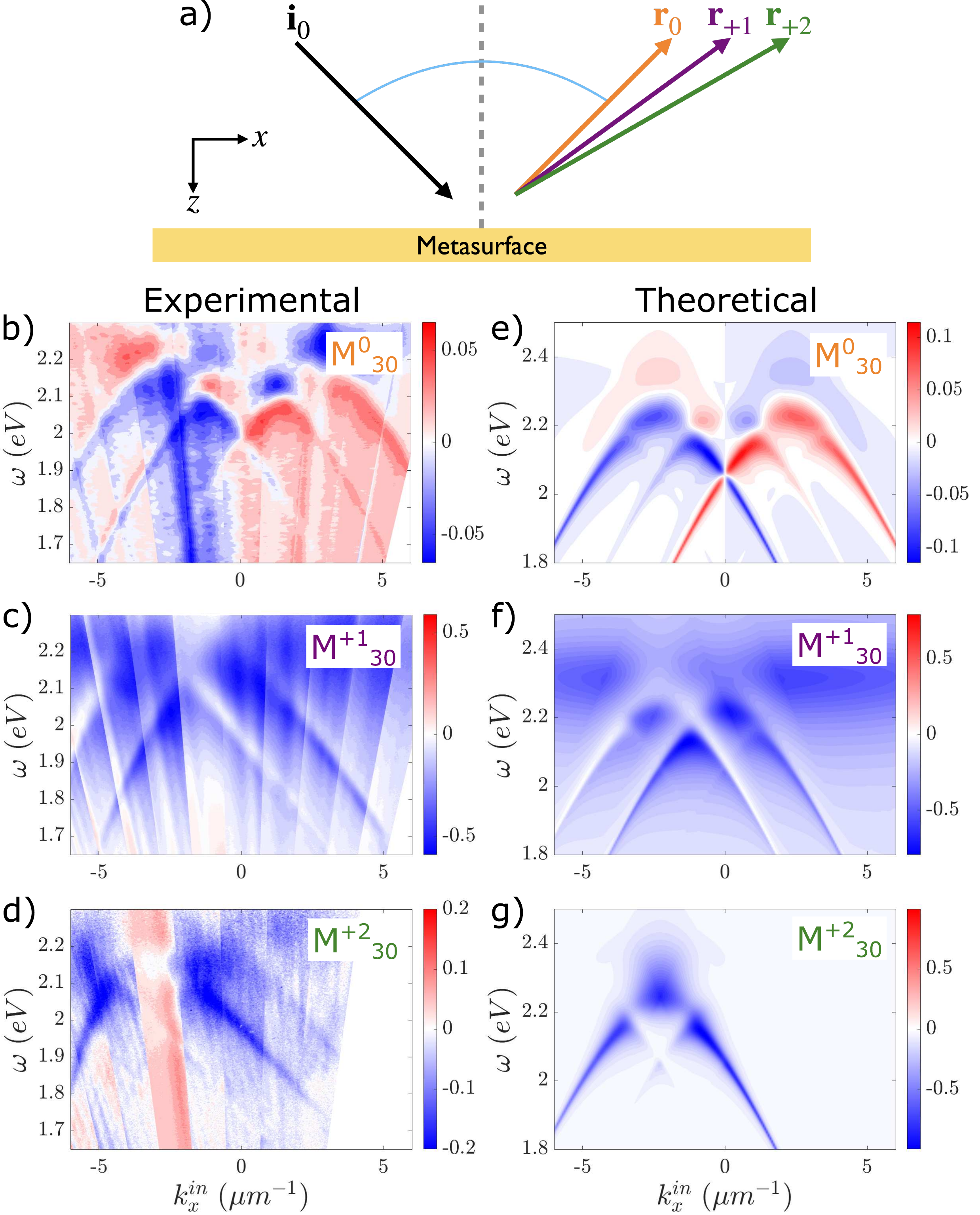}
	\caption{Mueller polarimetry for the system described in Fig.~\ref{fig:sample}(a). The grooves are filled with $SiO_2$, which also forms a 4 nm layer above the gold metasurface. 
	 (a) Scheme of the chosen Bragg modes. (b-d) Experimental $M_{30}^{n}(\omega, k_{x}^{in})$, for a reflected wave with $k_{x}^{out} = k_{x}^{in} + n k_{g}$.  (e-g) Theoretical $M_{30}^{n}(\omega, k_{x}^{in})$. The calculations have been performed within the SIBC approximation, and the groove dimensions have been phenomenologically enlarged by 1.25 times the skin depth to consider the field penetration in the metal~\cite{MartinMoreno2001}. }
	\label{fig:mueller}
\end{figure}

For a flat interface,  in-plane momentum is conserved, and $M$ can be measured for all $\omega$ and $k_x^{in}$ at once, using an objective whose back focal plane is fully illuminated with a collimated white light beam. However, the Mueller matrix can be defined and measured for each diffracted order $n$ in a periodically corrugated interface; see Fig.~\ref{fig:mueller}(a). Here we will analyze $M_{30}^{n}(\omega, k_{x}^{in})$, obtained from the polarization properties of the reflected light at Bragg orders $k_{x}^{out} = k_{x}^{in} + n k_{g}$, and restricting ourselves to $n=0, 1, 2$, for reasons explained below. 

The experimental results for the GPM shown in Fig.~\ref{fig:sample} are presented in Figures~\ref{fig:mueller}(b-d). 
Panels Figs.~\ref{fig:mueller}(e-g) show the numerical simulations performed with the CMM framework within the SIBC approximation.  
The calculations reproduce the main features found in the experiments. 
All panels show inverted parabolic features related to the resonant excitation of the SPP of the metasurface. 

Let us concentrate on the plasmonic resonances, which are the main reason for analyzing metallic GPMs. We find that \textit{resonant} reflection processes can be understood as two-step scattering processes. First, the incident light scatters with the surface picking up momentum $n_0G^{0} + n_{1} k_{g}$ and adding spin $-n_{1}$ (where $n_{1} = 0, \pm 1$).  If this scattered wave is an SPP, then it only has a $p$-polarization component.  A subsequent scattering of the SPP with the surface brings it into one of the considered diffraction orders (for that, the picked momentum must be $-n_0G^{0} + n_{2} k_{g}$, which adds spin $-n_{2}$).  Taking into account that the SPP excitation  can mathematically be represented as a projector onto the $p$-linear polarization, this two-step resonant scattering process results in the following  rules:
\begin{align}
    &k_x^{SPP}  = k_x^{in} + n_0 G^{0} + n_1 \, k_g, \label{eq:res_spp}\\
    &k_x^{out}  = k_x^{in} + (n_1 + n_2) \, k_g, \label{eq:res_mom}\\
    & \mathbf{\sigma}_{out}  = \sigma_{-n_{2}}\cdot \mathbf{p} \, (\mathbf{p}^{T} \cdot \sigma_{-n_{1}} \cdot \mathbf{i}_{0})
        \label{eq:res_spin}
\end{align}
where $\mathbf{p} = 2^{-1/2} \, (1,1)^{T}$ is the linear $p$-polarization and $\mathbf{i}_{0}$ the incident polarization (in the circular polarization basis), $\sigma_{\pm}$ are the Pauli matrices that increase and decrease spin, and we have written $\sigma_{0} \equiv \mathbb{1}$. These rules substitute the SML rules when SPPs are excited and are the ones to be used to understand and predict the polarization properties of resonant plasmonic structures.

We start by considering how this reasoning applies to the case $k_x^{out}  = k_x^{in} + k_g$, rendered in 
Figs.~\ref{fig:mueller}(c) and (f).  
The central parabola corresponds to the case $n=1, \, n_{1}=0, n_{2}=1$. As $n_{1}=0$, the unpolarized incident light maintains its spin after interacting with the surface and thus can excite the linearly polarized SPP.  The second scattering with the surface removes the $nG^{0}$ momentum. Still, it adds $+k_{g}$, thus decreasing the spin of the $p$-polarized wave, ending with spin $-$, which agrees with both experimental and computed results. 
The parabola that appears displaced to a smaller $k_x^{in}$ arises from the processes $n=1, \, n_1=1, \, n_2 = 0$.  Thus the incident photon has spin $-$ after the first scattering, still being able to excite the SPP.  As the second scattering with the grating has $n_{2}=0$, it conserves spin, and the reflected photon is linearly polarized (so, it has $M_{30} = 0$, in accordance with the experimental data). 
Thus, both parabolas involve resonant excitation of an SPP and two interactions with the holey array, described by one standard and one spin-orbit Bragg laws, being the difference in the order of the ``standard'' and ``geometric'' processes. 

Figs.~\ref{fig:mueller}(d,g) represent the case when the outcoming momentum is $+2k_g$ larger than the incident one. The existence of this diffraction order is a direct confirmation of the presence of the SML breakdown, as an exact SML would imply that the addition of $+2k_g$ momentum must be accompanied by a reduction of spin in -2 units, which is not possible (as $\sigma_{-}^{2} =0$ for spin 1/2 spinors). In this case, both experiment and calculation show only one plasmonic resonance, resulting in a spin $-$ in the reflected wave.  
The reason is that the two interactions with the array now involve spin-orbit Bragg laws. The first interaction of the incident wave with the holey surface and the resonant excitation of SPP is like in Figs.~\ref{fig:mueller}(c,f), but the second interaction also adds geometrical momentum and thus decreases spin, producing a reflected wave with spin $-$. 

Finally, Figs.~\ref{fig:mueller}(b,e) represent the case of specular reflection: $k_x^{out} = k_x^{in}$. In this case, on top of the 
processes described by Eqs.~\ref{eq:res_spp}-\ref{eq:res_spin} the direct reflection (the term $-\delta_{m0} \mathbf{i}_0$ in Eq.~\ref{eq:approxSML}) must also be considered, and it dominates the signal. For this reason, spin $-$ is fully reflected in the left parabola while spin $+$, which can couple to an SPP, is only partially reflected. Thus, overall, $M_{30}^{0}(\omega, k_{x}^{in})$ is negative in that spectral region, although much lower in magnitude than for the rest of Bragg orders ($n=1,2$) because of the dominance of the specular reflection.

Therefore, we have seen that the SPP acts as a filter to linear $p$-polarized light and also allows us to reach processes where the photon acquires an extra momentum $2k_g$, which would not be possible without the SML breakdown.


\section{Conclusion}
We have presented a rigorous theoretical analysis, based on a scattering formalism, that shows how spin-momentum locking in geometric phase metasurfaces emerges from the geometry (winding number) of the unit cell. Furthermore, we show that SML is an approximate symmetry with or without global lattice symmetries. SML breakdown terms yield couplings to Bragg orders that would otherwise be uncoupled, such as reflected waves that pick up two units of geometrical momentum. The origin of the SML breakdown is that each circularly polarized wave has an elliptical polarization when projected onto the surface (except for waves with momentum normal to the surface). This breakdown is particularly relevant when linearly polarized surface resonances (as surface plasmon polaritons) are excited. Our analysis shows how SML rules should be modified when surface modes are resonantly excited in the system. This process can be viewed as a two-step interaction with the metasurface, with the plasmon acting as a polarization filter. These modified SML rules perfectly agree with experimental results on the excitation of plasmonic resonances obtained with Mueller polarimetry.

Considering the crucial role played by spin-momentum locking in integrated quantum optical systems \cite{lodahl2017chiral}, our elucidation of the mechanism and its relation to the near-field will help understand and design plasmonic structures for the polarization control of light. This is important in the current applicative perspectives discussed currently in the context of optovalleytronic systems \cite{Li2021}, nonlinear hybrid metasurfaces \cite{Hu2019}, and topology-based high-resolution sensors \cite{Bozhevolnyi2017}.
 

\begin{acknowledgments}
We acknowledge Project No. PID2020-115221GB-C41, financed by MCIN/AEI/10.13039/501100011033, and the Aragon Government through Project Q-MAD. We acknowledge Project No. PID2020-115221GB-C41, financed by MCIN/AEI/10.13039/501100011033, and the Aragon Government through Project Q MAD. We also acknowledge the Interdisciplinary Thematic Institute QMat as part of the ITI 2021 2028 program of the University of Strasbourg, CNRS, and Inserm. This work was supported in part by IdEx Unistra (ANR 10 IDEX 0002), SFRI STRAT'US project (ANR 20 SFRI 0012), and by the University of Strasbourg Institute for Advanced Study (USIAS; ANR-10-IDEX-0002-02) under the framework of the French Investments for the Future Program.
\end{acknowledgments}



\end{document}


\title{Supplemental Material:\\Microscopic analysis of spin-momentum locking on a geometric phase metasurface}

\author{Fernando Lor\'en}
\affiliation{Instituto de Nanociencia y Materiales de Arag\'on (INMA), CSIC-Universidad de Zaragoza, 50009 Zaragoza, Spain\looseness=-1}
\affiliation{Departamento de F\'isica de la Materia Condensada, Universidad de Zaragoza, 50009 Zaragoza, Spain\looseness=-1}

\author{Gian L. Paravicini-Bagliani}
\affiliation{University of Strasbourg and CNRS, CESQ \& ISIS (UMR 7006), 8, all\'ee G. Monge, 67000 Strasbourg, France}

\author{Sudipta Saha}
\affiliation{University of Strasbourg and CNRS, CESQ \& ISIS (UMR 7006), 8, all\'ee G. Monge, 67000 Strasbourg, France}

\author{J\'er\^ome Gautier}
\affiliation{University of Strasbourg and CNRS, CESQ \& ISIS (UMR 7006), 8, all\'ee G. Monge, 67000 Strasbourg, France}

\author{Minghao Li}
\affiliation{University of Strasbourg and CNRS, CESQ \& ISIS (UMR 7006), 8, all\'ee G. Monge, 67000 Strasbourg, France}

\author{Cyriaque Genet} 
\email{genet@unistra.fr}
\affiliation{University of Strasbourg and CNRS, CESQ \& ISIS (UMR 7006), 8, all\'ee G. Monge, 67000 Strasbourg, France}

\author{Luis Mart\'in-Moreno}
\email{lmm@unizar.es} 
\affiliation{Instituto de Nanociencia y Materiales de Arag\'on (INMA), CSIC-Universidad de Zaragoza, 50009 Zaragoza, Spain\looseness=-1}
\affiliation{Departamento de F\'isica de la Materia Condensada, Universidad de Zaragoza, 50009 Zaragoza, Spain\looseness=-1}



\begin{abstract}
In this Supplemental Material, we develop the calculations underpinning the theoretical results presented in the main text, several verifications via simulations, and a description of the experimental procedure.
\end{abstract}

\maketitle
\tableofcontents


\section{1. Equations within the Coupled Mode Method for a real metal}\label{sec:real_eqs}
The  Coupled Mode Method (CMM) expands the EM fields in plane waves in the free space regions and waveguide modes inside the grooves and finds the field amplitudes by adequately matching the fields at the interfaces. The finite dielectric constant of a metal, $\epsilon_{M}$, can be considered within the surface impedance boundary conditions (SIBC). The limit $\epsilon_{M} \rightarrow - \infty$ corresponds to a perfect electric conductor (PEC), where the electromagnetic field does not penetrate the metal. The CMM equations are simpler in the PEC case and are considered in the main text when describing the spin-momentum locking. 

In this section, we present the more general equations used in the comparisons with the experiment to justify that the PEC limit already contains the physics described in the text, and then we particularize to the rectangular unit cell and the spatial distribution of holes considered in the main text.

The most general reciprocal lattice vectors are $\vec{G}_m = \vec{G}_{m_1,m_2} = m_1 \vec{G}_1 + m_2 \vec{G}_2$, where $\vec{G}_1,  \vec{G}_2$ are basis vectors for a considered lattice and where, to avoid cluttering the notation, we have englobed the pair of integers $(m_1,m_2)$ in the single index $m$.

In the past, the CMM equations have been expressed in terms of the modal amplitudes of the waveguide modes. This allows for a very compact representation, as a few waveguide modes are usually enough to obtain convergency.  The reflection coefficients are then obtained once the waveguide mode amplitudes have been computed.

Here we find it convenient to directly represent the CMM equations in terms of the reflection coefficients. Although more computationally demanding, this leads to a more transparent description of the different factors involved in the spin-momentum locking process. We thus take the expressions derived in \cite{deleon2008theory} and eliminate the waveguide mode amplitudes in favor of the reflection coefficients. 
By expressing the coefficients on the basis of circular polarizations, we obtain the following system of equations for the amplitudes of the reflected  Bragg modes.
\begin{equation}
  f_m^+ \, \mathbf{r}_{m} = - f_0^- \, \delta_{m 0} \, \mathbf{i}_0 +  C_{m 0} \, Y_0 \, \mathbf{i}_0 - \sum_{m'} C_{m m'}  \, Y_{m'} \, \mathbf{r}_{m'},
    \label{eq:SIBC}
\end{equation}
where $\mathbf{r}_{m}\equiv (r_m^+,r_m^-)^T$ is the spinor for the two circular polarizations for the reflected $m$-th Bragg mode with respect to its propagation direction (basis $k(m)$) and $\mathbf{i}_0$ is the amplitude of the incident wave in the circular basis with respect the incident direction ($k(0)$). $Y_{m}$ is a $2\times2$ matrix that depends on the modal admittances of the $m$-th Bragg mode. It is defined such that $Y_{m} = \bar{Y}_{m} \, \mathbb{1} + \Delta_{m} \, \sigma_x$, where $\mathbb{1}$ is the $2\times 2$ identity matrix and  $\sigma_x =\begin{pmatrix}0 & 1 \\ 1&0\end{pmatrix} $ is a Pauli matrix acting in circular polarization space.  
The coefficient $\bar{Y}_{m} = (Y_{mp}+Y_{ms})/2$ is the average of the modal admittances in the linear $p-s$ polarized basis, while $\Delta_{m} = (Y_{mp}-Y_{ms})/2$.
For a plane wave with frequency $\omega$ and in-plane wavevector $k_{m}$ propagating in a uniform medium with dielectric constant $\epsilon$, $Y_{mp} = \epsilon/q_{mz}$ and $Y_{ms} = q_{mz}$, where $q_{mz} = \sqrt{\epsilon -  (c k_{m}/ \omega)^{2}}$ (c being the speed of light).  

The coefficients $f_m^{\pm}$ are $2\times2$ matrices in the circular polarization basis with respect to the propagation direction of the $m$-th Bragg mode ($k(m)$):
\begin{equation}
	f_m^{\pm} = \frac{1}{2} \begin{pmatrix}
	f_{mp}^{\pm}+f_{ms}^{\pm} & f_{mp}^{\pm}-f_{ms}^{\pm}\\
	f_{mp}^{\pm}-f_{ms}^{\pm} & f_{mp}^{\pm}+f_{ms}^{\pm}\\
	\end{pmatrix},
\end{equation}
with $f_{m\sigma}^{\pm} = 1 \pm z_s Y_{m\sigma}$, where $\sigma = \left\{ s, p \right\}$ and $z_{s} \equiv 1/\sqrt{\epsilon_{M}}$ is the surface impedance. 
In fact, in this work, we take $z_{s} \equiv 1/\sqrt{\epsilon_{M} +1}$, which is a phenomenological correction that provides the exact dispersion relation of SPPs in a metal-vacuum interface. 

The expression for the geometric couplings between Bragg modes $C_{m m'}$, in the circular polarization basis with respect to the propagation direction, is also a $2\times2$ matrix acting on polarization space. It is easier to compute on the linear polarization $p-s$ basis with respect to the $\vec{u}_z$ direction ($C^{Lz}_{mm'}$). In this case, within the SIBC and considering only the fundamental waveguide mode inside the rectangular grooves, the components of the $C^{Lz}_{mm'}$ matrix are:
\begin{equation}
	C_{m\sigma m'\sigma'}^{Lz} = \underbrace{\frac{1}{Y}}_\textrm{\textcolor{red}{impedance}} 
	\cdot \underbrace{\frac{f^- \, (1+\Phi)}{1-\Phi \, f^- / f^+}}_\textrm{\textcolor{red}{Fabry-Perot}} \cdot \underbrace{\sum_{\alpha=0}^{N-1} S_{m\sigma \alpha} S_{m' \sigma' \alpha}^*}_{\textrm{\textcolor{red}{geometry}}},
	\label{eq:geometric}
\end{equation}

where $\Phi = - e^{i2 k_z^w d}$ and $k_{z}^{w}$ is the propagation constant along the z-direction of the fundamental waveguide mode.  For a rectangular hole with long side $b$, filled with a material with dielectric constant $\epsilon_{d}$,  $k_z^w = \sqrt{\epsilon_{d} (\omega/c)^{2} - k_w^2}$,  with $k_w = \pi / b$. The admitance of the fundamental mode is $Y = k_z^w c / \omega$, and $f^{\pm} = 1 \pm z_{s} Y$.

As highlighted in Eq.~\ref{eq:geometric}, $C^{Lz}_{mm'}$ has three factors. The first term is the impedance of the fundamental waveguide mode and is responsible for the potential existence of single-hole resonances associated with the cutoff of the waveguide mode.  The second term is related to the bouncing back and forth of EM fields inside the groove, and it is responsible for the potential existence of Fabry-Perot resonances inside the groove. The third term contains the overlaps between the Bragg and waveguide modes in all the apertures, and it is where the geometrical arrangements of holes in the unit cell are encoded in the CMM equations. %

The overlapping integral between the $m$-th Bragg mode and the waveguide mode of the $\alpha$-th hole is
\begin{equation}
	S_{m\sigma \alpha} = \int_{\text{hole}} \vec{E}_{m\sigma}^{\dagger}(\vec{r}) \cdot \vec{E}_{\alpha}(\vec{r}) d\vec{r},
	\label{eq:overlap_general}
\end{equation}
where $\vec{E}_{m\sigma}(\vec{r})$ and $\vec{E}_{\alpha}(\vec{r})$ are the in-plane electric field of the $m$-th Bragg mode with linear polarization $\sigma$ ($s$ or $p$) and fundamental mode of the $\alpha$-th aperture, respectively,   
and $\dagger$ denotes conjugate transpose. The integral is over the aperture opening. 

The expression for EM plane waves is $\vec{E}_{m\sigma}(\vec{r}) = \vec{u}_{m\sigma} e^{i \vec{k}_m \vec{r}} / \sqrt{A}$, where $A$ is the area of the unit cell, $\vec{k}_m =\vec{k}^{in} + \vec{G}_{m} = k_x^m \vec{u}_x + k_y^m \vec{u}_y$ and $\vec{u}_{m \sigma}$ is an unitary bivector that depends on the polarization. In general, $\vec{u}_{mp} = (k_x^m \vec{u}_x + k_y^m \vec{u}_y) / k^m$ and  $\vec{u}_{m s} = (-k_y^m \vec{u}_x + k_x^m \vec{u}_y) / k^m$, where $k_{m} = \sqrt{(k_x^m)^{2} +(k_y^m)^{2}}$. 

The expression for $\vec{E}_{\alpha}(\vec{r})$ is best expressed in a system of coordinates that has the origin at the center of the $\alpha$-th hole, and it is rotated by an angle $\theta_{\alpha}$.  This is, by writing $\vec{r} = \vec{r}_{\alpha} + R^{-1}(\theta_{\alpha})\vec{r}\,'$,  with $\vec{r}_{\alpha}$ marking the position of the $\alpha$-th aperture centre and:
\begin{equation}
\begin{pmatrix}
x'\\
y'
\end{pmatrix} = 
\underbrace{\begin{pmatrix}
\cos\theta_{\alpha} & \sin\theta_{\alpha}\\
-\sin\theta_{\alpha} & \cos\theta_{\alpha}
\end{pmatrix}}_{R(\theta_\alpha)}
\begin{pmatrix}
x - x_{\alpha}\\
y- y_{\alpha}
\end{pmatrix}, \qquad
\begin{pmatrix}
x- x_{\alpha}\\
y- y_{\alpha}
\end{pmatrix} = 
\underbrace{\begin{pmatrix}
\cos\theta_{\alpha} & -\sin\theta_{\alpha}\\
\sin\theta_{\alpha} & \cos\theta_{\alpha}
\end{pmatrix}}_{R^{-1}(\theta_\alpha)}
\begin{pmatrix}
x'\\
y'
\end{pmatrix}.
\end{equation}

With this,  $\vec{E}_{\alpha}(\vec{r}) = \sqrt{\frac{2}{ab}} \sin \left(k_w (y' + b/2)\right) \vec{u}_x\!'$ and the overlapping integral reads
\begin{eqnarray}
	S_{m \sigma \alpha} &=& \int_{\text{hole}} \vec{u}_{\sigma}^T \frac{1}{\sqrt{A}} e^{-i \vec{k}_m \vec{r}} \,  \sqrt{\frac{2}{ab}} \sin \left(k_w (y' + b/2)\right) \vec{u}_x\!'  d\vec{r}\,' \\
          &=& \sqrt{\frac{2}{Aab}} v_{m\sigma \alpha} e^{-i \vec{k}_m \vec{r}_{\alpha}}\int_{\text{hole}}  e^{-i \vec{k}_m R^{-1}(\theta_{\alpha})\vec{r}\,'}\sin \left(k_w (y' + b/2)\right)  d\vec{r}\,'.
\end{eqnarray}

where the integral is over the area $\{x' \in ( -a/2, a/2), y' \in ( -b/2, b/2)\}$, and we have defined  $v_{m\sigma \alpha} = \vec{u}_{\sigma}^T R^{-1} (\theta_{\alpha}) \vec{u}'_x$. 

The integral is straightforwardly solved, giving:
\begin{equation}
	S_{m \sigma \alpha} = \sqrt{\frac{a b}{2 A}} \, v_{m\sigma \alpha} \, e^{-i \vec{k}_m \vec{r}_{\alpha}} \, \text{sinc}(k_x'^{m} a /2) \, (\text{sinc}((k_y'^{m} + k_w) b / 2) + \text{sinc}((k_y'^{m} - k_w) b / 2) ),
	\label{eq:overlaps}
\end{equation}
where $\text{sinc}(x) \equiv \frac{\sin{(x)}}{x}$ is the cardinal sine function, $\vec{k}'_m = \vec{k}_m R^{-1} (\theta_{\alpha}) = k_{x}'^{m} \vec{u}'_{x} + k_{y}'^m \vec{u}'_y = k_x^m \cos \theta_{\alpha} \vec{u}'_x - k_x^m \sin \theta_{\alpha} \vec{u}'_y$,  $v_{m p \alpha} = (k_x^m \cos{\theta_{\alpha}} + k_y^m \sin{\theta_{\alpha}}) / k^m$ and $v_{m s \alpha} = (-k_y^m \cos{\theta_{\alpha}} + k_x^m \sin{\theta_{\alpha}}) / k^m$.

Notice that, for very small grooves,  the dependence of the cardinal sine on wavevector disappears: $\text{lim}_{a \rightarrow 0} \text{sinc}(k_x'^m a /2) = 1$ and $\text{lim}_{b \rightarrow 0} (\text{sinc}(k_{y}'^m + k_w) b / 2) + \text{sinc}(k_y'^m - k_w) b / 2) ) = 4/\pi$.  We call this the small-hole limit,  where the overlapping integral becomes
\begin{equation}
	S_{m \sigma \alpha} = \sqrt{\frac{a b}{2 A}} \frac{4}{\pi} \, v_{m \sigma \alpha} \, e^{-i \vec{k}_m \vec{r}_{\alpha}}.
	\label{eq:overlaps-small}
\end{equation}

The small-hole limit allows an analytical expression for the geometric couplings and leads to the spin-momentum locking reflected in Eq. (2) of the main text.  
For large enough grooves,  the momentum dependence of the cardinal sines is relevant and must be considered.  This, in principle,  originates another mechanism for the breakdown of spin-momentum locking.  
In practice, this mechanism creates a small effect, as seen in the comparison between experimental Mueller matrices and the theoretical ones, which we compute considering the full expression in Eq.~\ref{eq:overlaps} and applying it to the system considered in the main text (see Eq.~\ref{eq:overlaps_ms}).

With this, we can compute $C^{Lz}_{mm'}$ in the linear $p-s$ polarization basis with respect to the $\vec{u}_z$ direction. In the small-hole limit:

\begin{equation}
	C_{mm'}^{L_z} = \frac{1}{Y} \, \frac{f^+ f^- (1 +\Phi)}{f^+ - f^- \Phi} \, \frac{8ab}{\pi^2 A}\sum_{\alpha=0}^{N-1}
	e^{i (\vec{k}_{m'} - \vec{k}_{m}) \vec{r}_{\alpha}} \begin{pmatrix}
	v_{m p \alpha} v_{m' p \alpha} & v_{m p \alpha} v_{m' s \alpha} \\
	v_{m s \alpha} v_{m' p \alpha} & v_{m s \alpha} v_{m' s \alpha}\\
	\end{pmatrix}.
\end{equation}

In the {\it circular polarization} basis with respect to the $\vec{u}_z$ direction, $C_{mm'}^{z}$ becomes:

\begin{eqnarray}
	C_{mm'}^{z} &=& \frac{1}{Y} \, \frac{f^+ f^- (1 +\Phi)}{f^+ - f^- \Phi}  \, \frac{4ab}{\pi^2 A} \sum_{\alpha=0}^{N-1} e^{i (\vec{k}_{m'} - \vec{k}_{m}) \vec{r}_{\alpha}}
	 \begin{pmatrix}
	c_{++} & c_{+-} \, e^{- i 2 \theta_{\alpha}} \\
	c_{-+} \, e^{ i 2 \theta_{\alpha}} & c_{--} \\
	\end{pmatrix} ,
		\label{eq:SMLgeneral}
\end{eqnarray}

where we have defined $c_{s s' } = \braket{\vec{k}_m | \vec{s}} \braket{\vec{s}' | \vec{k}_{m'}} / (k^m k^{m'})$, being $\vec{s} = \vec{u}_x + i s \vec{u}_y$, with $s =  \pm $. These $c_{ss'}$ are the projections of the Bragg modes $m$ and $m'$ with the circular polarizations $s$ and $s'$, respectively.


As said, the CMM equations for a PEC can be readily obtained by taking $\epsilon_M \rightarrow - \infty$. In this case, $z_s \rightarrow 0$, so $ f^{\pm}  \rightarrow 1$ and $ f_{m}^{\pm} \rightarrow \mathbb{1} $. 



\begin{figure}[ht]
\includegraphics[width=0.6\columnwidth]{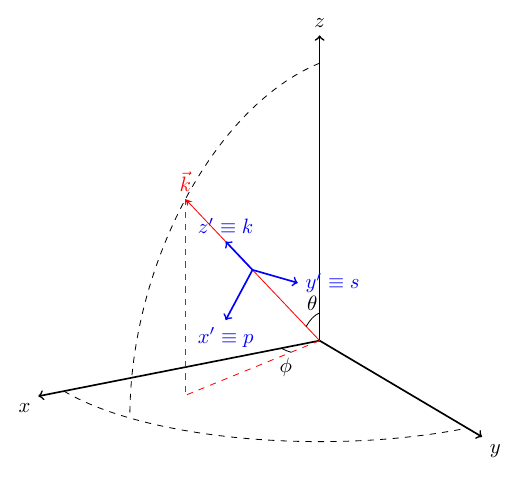}
\caption{Scheme of the standard reference framework $xyz$ and the propagation direction framework $x'y'z'$, where $x' \equiv p$, $y' \equiv s$ and $z' \equiv k$, being $k$ the wavevector.}
\label{fig:kps_basis}
\end{figure}

However, in Eq.~\ref{eq:SIBC}, each reflection coefficient (and the rest of the terms) is referred to the circular polarization basis defined with respect to its own propagation direction. 
These differences between the basis set defined with respect to $\vec{u}_{z}$ and that defined with respect to propagation direction can be seen in Fig.~\ref{fig:kps_basis}.  This figure shows that, in the new basis, the linear polarizations $p$ and $s$ in the rotated basis are different from the original ones ($p=x$ and $s=y$).

To transform from one circular basis to the other,  we define the $2 \times 2$ matrix $R^{z \leftarrow k(m)}$ such that,  given the reflection coefficient $\mathbf{r}_{m}$ (defined in the circular basis normal to its propagation direction),  $R^{z \leftarrow k(m)} \mathbf{r}_{m}$ returns these spinor components in the circular basis defined in the plane normal to $\vec{u}_{z}$. The matrix $R^{k(m)\leftarrow z} = \left(R^{z \leftarrow k(m)} \right)^{-1}$ does the opposite transformation.

In terms of these matrices $R$, the geometric couplings are expressed as:
\begin{equation}
    C_{mm'} = R^{k(m)\leftarrow z} \, C_{mm'}^{z} \, R^{z \leftarrow k(m')},
\end{equation}
Recall that $C_{mm'}$ connects a spinors $\mathbf{r}_{m'}$ and $\mathbf{r}_{m}$, defined in different circular basis (described with respect to the propagation direction of the $m'$-th an $m$-th Bragg mode, respectively).  


We find that $R^{k(m)\leftarrow z} = \frac{1}{2} \left[ (\sqrt{q_{mz}^2 + q_m^2}/q_{mz} + 1) \, \mathbb{1} + (\sqrt{q_{mz}^2 + q_m^2}/q_{mz} - 1) \, \sigma_x \right]$, where $q_m = c k_m / \omega$. So terms proportional to $\sigma_x$ appear in $C_{mm'}$,  mixing the circular polarization states. As mentioned in the main text, this is another source of spin-momentum locking breakdown (in addition to the one that appears from the differences between modal admittances provide).

\subsection{Particular case: rectangular lattice with spatially rotated grooves}
Here, we evaluate the expressions derived in the previous section for the particular metasurface considered the main text.  In this case, the unit cell is rectangular,  with reciprocal lattice vectors that can be written as $\vec{G}_{m}  = \vec{G}_{m,n_{y}}  = 2\pi m / (NL) \, \vec{u}_x + 2\pi n_y / L \, \vec{u}_y$.  Notice that, to shorten the notation, we have englobed in the single index $m$ the pair $(m, n_{y})$. 

As the next section demonstrates, a very good approximation can be obtained by considering only Bragg modes with $n_{y}=0$.  Within this approximation, $m$ is just the index of the Bragg mode in the $\vec{u}_x$ direction. If, additionally, the incident wave has $k_{y}=0$, the unitary vectors describing the in-plane component of the electric field are  $\vec{u}_{mp} = \vec{u}_x$ and $\vec{u}_{ms} = \vec{u}_y$.  In this case, $v_{m p \alpha} = \cos\theta_{\alpha}$ and $v_{ms\alpha} = \sin\theta_{\alpha}$ (notice that $v_{m p \alpha}$ and $v_{m s \alpha}$ do not depend on the Bragg mode $m$ any longer). 

The general expression for the overlapping integrals is:
\begin{equation}
	S_{m \sigma \alpha} = \sqrt{\frac{a b}{2 A}} \, v_{m\sigma \alpha} \, e^{-i k_x^m x_{\alpha}} \, \text{sinc}(k_x'^{m} a /2) \, (\text{sinc}((k_y'^{m} + k_w) b / 2) + \text{sinc}((k_y'^{m} - k_w) b / 2) ),
	\label{eq:overlaps_ms}
\end{equation}
and in the small-hole approximation:
\begin{equation}
	S_{m \sigma \alpha} = \frac{1}{\pi} \sqrt{\frac{a b}{A}} \, v_{m \sigma \alpha} \, e^{-i k_x^m x_{\alpha}}.
\end{equation}

With this, we can compute $C^{Lz}_{mm'}$ in the linear $p-s$ polarization basis with respect to the $\vec{u}_z$ direction. In the small-hole limit and for $k_{y}^{m} = k_{y}^{m'} =0$:

\begin{equation}
	C_{mm'}^{L_z} =\frac{8}{\pi^2} \, \frac{1}{Y} \, \frac{f^+ f^- \, (1+\Phi)}{f^+-\Phi \, f^- }\,  \frac{ab}{A}\sum_{\alpha=0}^{N-1}
	e^{i (k_x^{m'} - k_x^{m}) x_{\alpha}} \begin{pmatrix}
	\cos^2\theta_{\alpha} & \cos\theta_{\alpha} \sin\theta_{\alpha} \\
	\cos\theta_{\alpha} \sin\theta_{\alpha} & \sin^2\theta_{\alpha}\\
	\end{pmatrix}.
\end{equation}

In the {\it circular polarization} basis with respect to the $\vec{u}_z$ direction, $C_{mm'}^{z}$ becomes:

\begin{eqnarray}
	C_{mm'}^{z} &=& \frac{4}{\pi^2} \, \frac{1}{Y} \, \frac{f^+ f^- \, (1+\Phi)}{f^+-\Phi \, f^- }\,  \frac{ab}{A} \sum_{\alpha=0}^{N-1}
	 \begin{pmatrix}
	e^{i (k_x^{m'} - k_x^{m}) x_{\alpha}} & e^{i (k_x^{m'} - k_x^{m}) x_{\alpha} - i 2 \theta_{\alpha}} \\
	e^{i (k_x^{m'} - k_x^{m}) x_{\alpha}+ i 2 \theta_{\alpha}} & e^{i (k_x^{m'} - k_x^{m}) x_{\alpha}} \\
	\end{pmatrix} \\
	&=&  C \left( \delta_{m, m'+n_0 N} \, \mathbb{1} + \sum_{s=\pm} \delta_{m, m'+ n_0 N-2 n_w s} \, \sigma_s \right),
	\label{eq:SML}
\end{eqnarray}
where we have used that $x_{\alpha} = \alpha\, L$ and $\theta_{\alpha} = 2 \pi \alpha n_{w} / N$.  The circular polarization is described by the index  $s = \pm$,  $\mathbb{1}$ is the $2\times2$ identity matrix, and $\sigma_{\pm}$ are Pauli Matrices: $\sigma_{-} = \begin{pmatrix}0 & 0 \\ 1&0\end{pmatrix}$, $\sigma_{+} = \begin{pmatrix}0 & 1 \\ 0&0\end{pmatrix}$.
Finally, the proportionality constant in the coupling coefficients $C_{mm'}^{z}$ is
\begin{equation}
C =  \frac{4}{\pi^2} N \, \frac{1}{Y} \, \frac{f^+ f^- \, (1+\Phi)}{f^+-\Phi \, f^- }\,  \frac{ab}{A} .
\end{equation}


\section{2. Justification of the minimal model}

In this section, we will study how the number of Bragg modes affects the results of our simulations, and find the minimal number of them required to encapsulate all the physics. For that, we first show that considering only Bragg modes with $n_{y}=0$ provides a good approximation.  And, second, that just considering a few Bragg modes in the $\vec{u}_{x}$ direction is enough. 


\subsection{Reduction in the number of reciprocal lattice vectors along the $y$-direction}
\begin{figure}[ht]
\includegraphics[width=0.5\columnwidth]{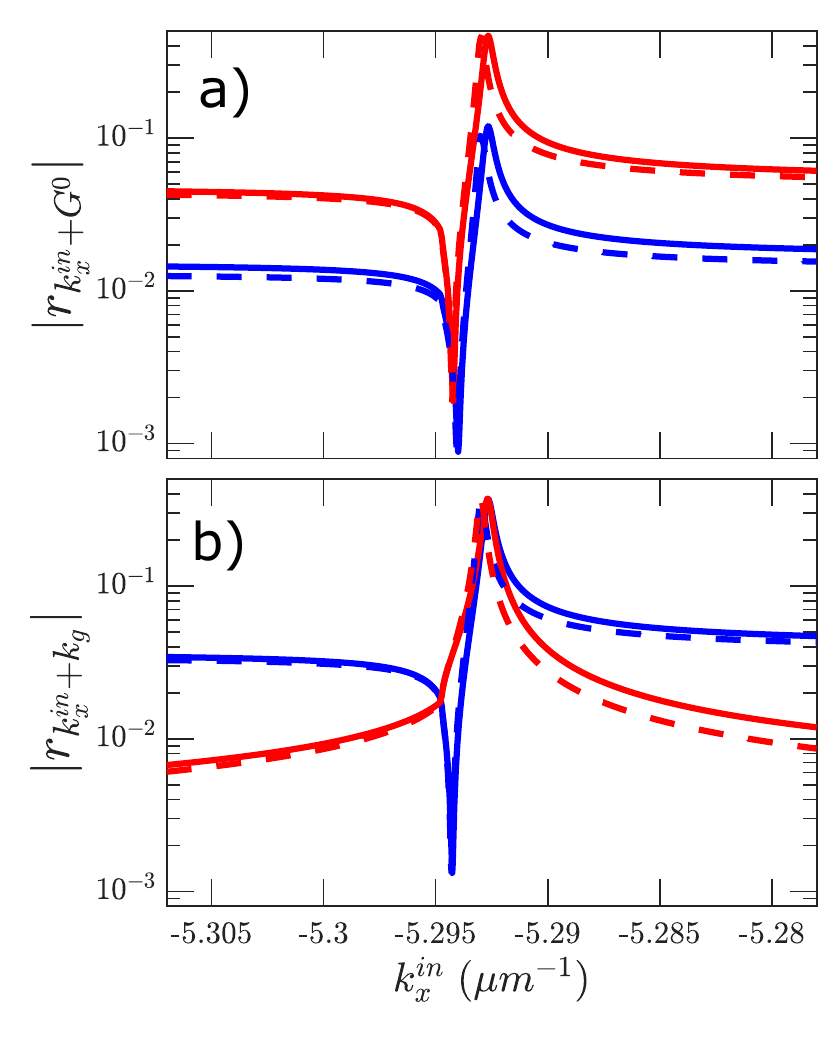}
\caption{Absolute value of the spin components of two reflection coefficients in terms of the incident momentum $k_x^{in}$. 
Panel (a): $r_{(k_x^{in}+G^0,0)}$. Panel (b): $r_{(k_x^{in}+k_g,0)}$.  Red: spin $+$ component. Blue: spin $-$ component. The solid lines render the calculations considering only Bragg orders with $n_y=0$, while the dashed include $n_y = \{-1, 0, 1\}$. 
Chosen values for the computation parameters: $\omega = 2.1 \, eV$, $a = 80 \, nm$, $b = 220 \, nm$, $d = 60 \, nm$, $L = 460 \, nm$, $n_w = 1/2$ and $N = 6$, based on the experimental shown in the main text. This computation has been done with the PEC approximation.}
\label{fig:rs_y}
\end{figure}

In the main text, we showed results on a minimal model that considered only reciprocal lattice vectors along the $x$-direction (the one along which the rotation of grooves occurs). In that case,  the integers $m$ and $m'$  label the reciprocal lattice vector $\vec{G}_{m} = 2 \pi m / (NL) \vec{u}_x$. 
However, in general, Bragg modes with components along both $\vec{u}_x$ and $\vec{u}_y$ directions must be considered. 
We do that in this section, showing that the approximation taken to produce the minimal model does not alter the conclusions reached in the paper. 

In the more general case, the momentum of the incident plane wave is $\vec{k}_{in} = k_x^{in} \, \vec{u}_x + k_y^{in} \, \vec{u}_y$.  For the rectangular lattice considered in this work, Bragg modes are defined by reciprocal lattice vectors characterized by two integers $m$ and $n_y$, such that $\vec{G}  = 2\pi m / (NL) \, \vec{u}_x + 2\pi n_y / L \, \vec{u}_y$. 

The geometric couplings now depend on both $m$ and $n_y$ and the corresponding $m'$ and $n_y'$. This is, we have $C_{(mn_y)(m' n_y')}$ instead of the previous $C_{mm'}$.  Performing the calculations also considering modes in the $\vec{u}_y$ direction, the geometric couplings in the circular basis with respect to the direction $\vec{u}_z$ read, in the limit of small holes, 
\begin{equation}
    C_{(mn_y)(m' n_y')}^{z} = C \left( \delta_{m, m'+n_0 N} \, \mathbb{1} + \sum_{s=\pm} \delta_{m, m'+ n N-2 n_w s} \, \sigma_s \right) \delta_{n_y, n_y' + n_0'} \, 
\end{equation}
where $n_0$ and $n_0'$ are integers.

The inclusion of reciprocal vectors along the $y$-direction opens new diffraction channels. However, as expected as inversion symmetry is maintained along the $y$-direction, it does not modify the spin-orbit conditions. 

To visualize how the inclusion of reciprocal vectors along the $y$-direction influences the results, we show in Fig.~\ref{fig:rs_y} the same reflection coefficients as those rendered in Fig.~2 of the main text.  The results considering only reciprocal lattice vectors with $n_{y}=0$ are shown by the solid line, while those with $n_{y}= 0, \pm1$ are given by the dashed lines (in both cases, $m$ ranges from -7 to 7).  
This figure shows that including more modes slightly affects the peak positions (displacing the plasmonic peaks to lower momenta), but the physics remains the same. 
Thus, considering only modes with $n_{y}=0$ is fully justified.

\subsection{Reduction in the number of reciprocal lattice vectors along the $x$-direction}
Similarly,  although, in principle, an infinite number of reciprocal vectors along the $x$-direction should be included,  the reflection coefficients can be obtained considering a highly reduced number of them. A minimal model for studying the spin-momentum locking in the considered system only requires $|m| \le   N+2 \, n_w$. 
\begin{figure}[ht]
\includegraphics[width=\columnwidth]{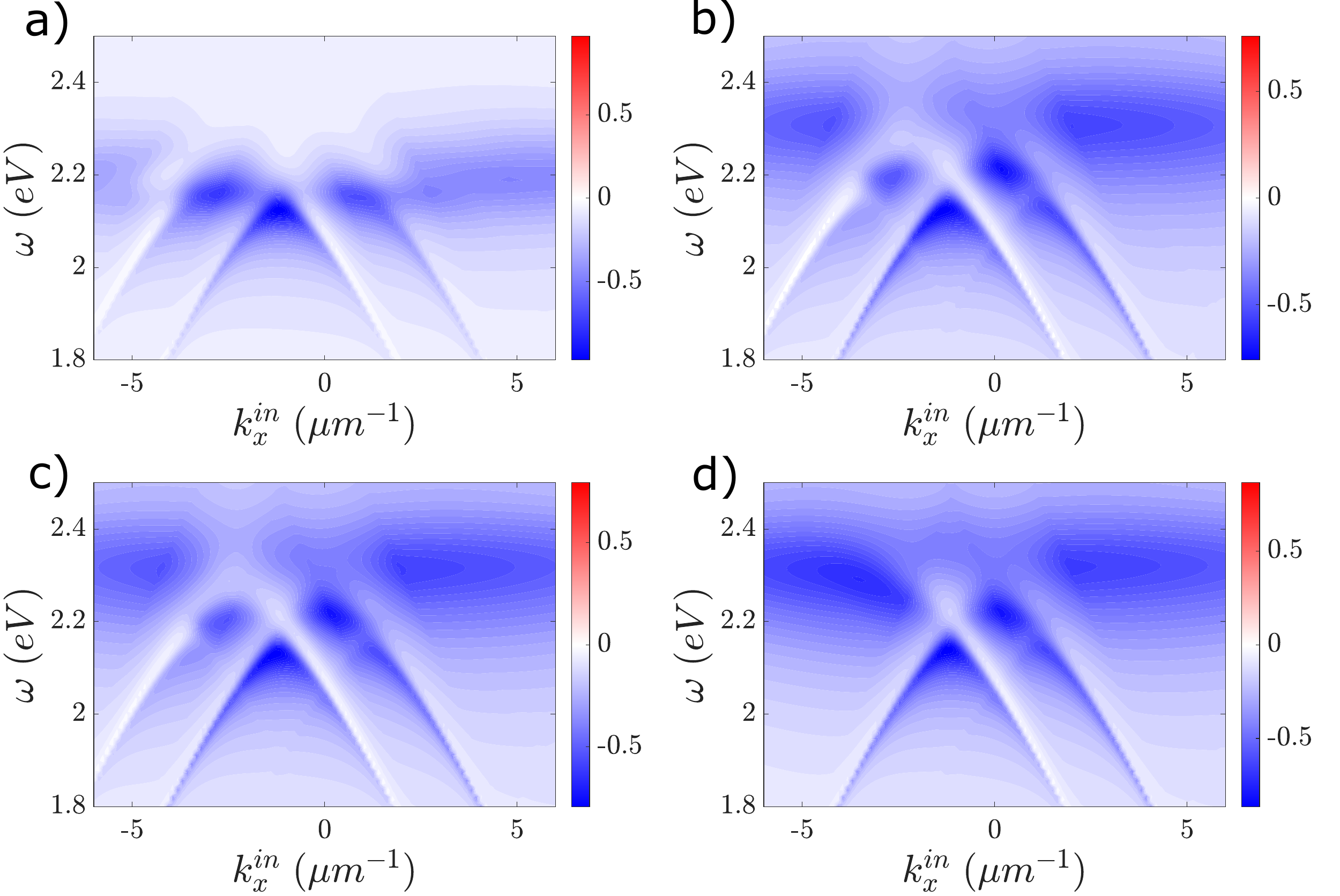}
\caption{$M_{30}^{+1}$ in terms of the incident photon energy $\omega$ and the incident momentum $k_x^{in}$.  $M_{30}$ provides the imbalance between energy radiated with spin $+$ and spin $-$ when the incident light is not polarized.  $M_{30}^{+1}$ is the matrix element  $M_{30}$ for the reflected wave with $k_x^{out} = k_x^{in} + k_{g}$.
Blue indicates a spin $-$ imbalance, while red signals that reflection with spin $+$ dominates. 
(a) $M = 6N, \, N_y = N$. (b) $M = 6N, \, N_y = 0$. (c) $M = N+2 \, n_w, \, N_y = 0$. (d) $M = N+2 \, n_w-1, \, N_y = 0$.  The calculations are for the system depicted in Fig. 1 of the main text, with the following parameters: $a = 80 $~nm, $b = 220 $~nm, groove depth $d = 60 $~nm, $L = 460 $~nm. The grooves are filled with $Si O_2$, which also forms a $4$~nm layer above the gold metasurface. The calculations have been performed within the SIBC approximation, and the groove dimensions have been phenomenologically enlarged by 1.25 times the skin depth to consider the field penetration in the metal.}
\label{fig:fewXmodes}
\end{figure}

To visualize how the inclusion of reciprocal vectors along the $x$-direction influences the results, we represent the component $M_{30}^{+1}$ (see Fig.~\ref{fig:fewXmodes}) of the Mueller matrix, for the case of $k_x^{out} = k_{x}^{in} + k_g$, and consider a different number of modes reciprocal lattice vectors (defined by $\vec{G}  = 2\pi m / (NL) \, \vec{u}_x + 2\pi n_y / L \, \vec{u}_y $) by setting values for $M, \, N_{y}$ such that $|m| \le M$ and $|n_{y}| \le N_{y}$.

In Fig~\ref{fig:fewXmodes}(a), a large number of reciprocal lattice vectors are considered in both $x$ and $y$ directions. Inverted parabolas, related to the resonant excitation of SPPs in the structure, are visible in $M_{30}^{+1}$.  The central parabola is blue (reflection mainly with spin $-$ polarization), while the one displaced to smaller wavevectors appears in white, meaning that the reflected wave is linearly polarized. Fig.~\ref{fig:fewXmodes}(b) is computed considering only reciprocal lattice vectors along the $x$-direction ($N_{y}=0$), while Fig.~\ref{fig:fewXmodes}(c) additionally reduces the number of reciprocal lattice vectors along the $x$-direction to those in the minimal model.  Both panels, Fig.~\ref{fig:fewXmodes}(b) and (c), show the same structures as Fig.~\ref{fig:fewXmodes}(a), justifying the minimal model used in the main text. That this model is minimal is shown in Fig.~\ref{fig:fewXmodes}(d), where $M$ has been further reduced to $M = N + 2 \, n_w - 1$.  As seen in the figure, this value of $M$ fails to capture one branch of the white plasmonic parabola seen in all other panels. 

This behavior has also been checked for other components of the Mueller matrices and other reflection orders. 


\section{3. Validity of the small-hole approximation}
As we have discussed before, when the apertures are much smaller than the wavelength, the expression in the overlaps in Eq.~\ref{eq:overlaps} can be simplified, leading to Eq.~\ref{eq:overlaps-small}, from where we can derive the spin-momentum locking that appears from the geometric couplings. Physically, the small-hole approximation considers the aperture's interaction with a spatially constant field over it; this is, it treats the hole as a dipole. 

\begin{figure}[ht]
\includegraphics[width=0.5\columnwidth]{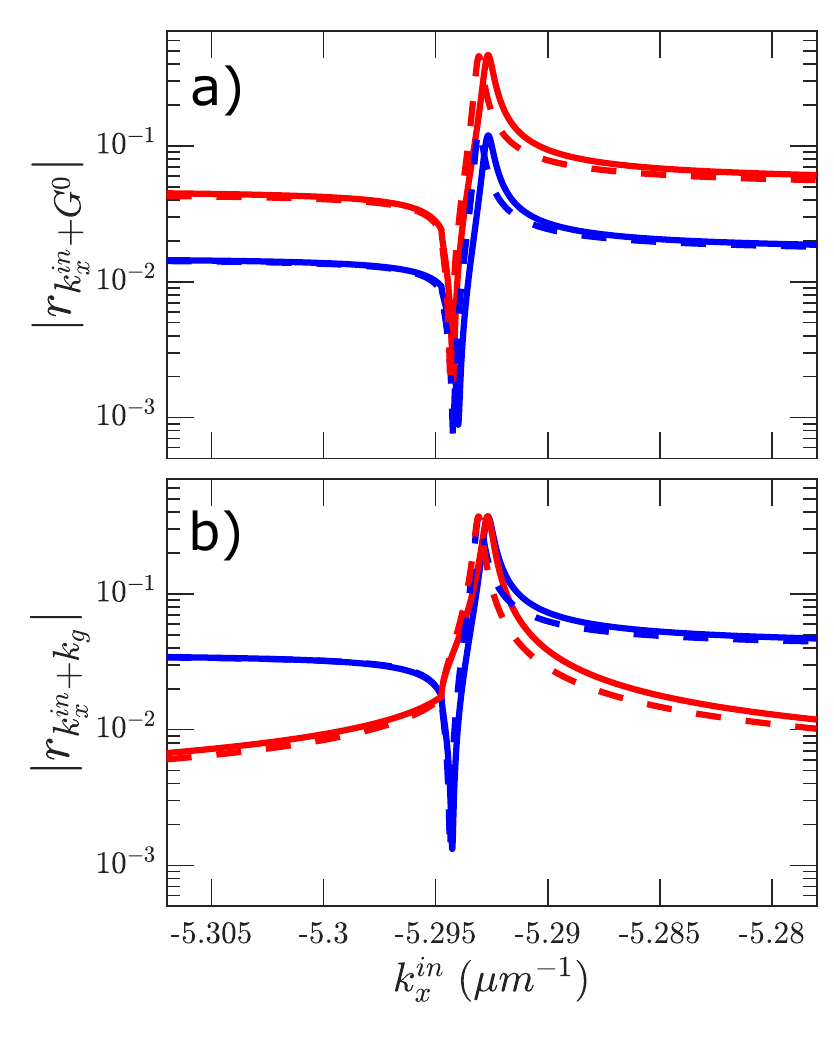}
\caption{Absolute value of the spin components of the reflection coefficient as a function of the incident momentum $k_x^{in}$.  Two diffraction orders are considered: (a) $r_{k_x^{in}+G^0}$, (b) $r_{k_x^{in}+k_g}$.  The spin $+$ component is represented by the red line, while the spin $-$ component is shown by the blue line.  The solid line is for the small-hole approximation, while the dashed line shows the exact result. Chosen values for the computation: $\omega = 2.1 \, eV$,  $a = 80 \, nm$, $b = 220 \, nm$, $d = 60 \, nm$, $L = 460 \, nm$, $n_w = 1/2$ and $N = 6$, based on the experimental shown in the main text. This computation has been done in the PEC approximation for spin-momentum locking breakdown.}
\label{fig:rs_PEC_comparison}
\end{figure}

For realistic values of the groove geometry, we find that the small-hole approximation is an excellent one. This is illustrated in Fig.~\ref{fig:rs_PEC_comparison} here; we present the reflection amplitude for two different Bragg orders ($r_{k_x^{in}+G^0}$ and $r_{k_x^{in}+k_g}$) and the two circular polarizations (in red, we represent the reflection coefficient with the same circular polarization as the incoming wave and in blue for the opposite polarization).  The solid lines are the same that we presented in the main text, i.e., within the small-hole approximation. The two other dashed curves correspond to the computation considering the exact overlap. Clearly, the small-hole approximation is an excellent one. 

\section{4. Mueller experimental set-up}
\begin{figure}[ht]
\includegraphics[width=\columnwidth]{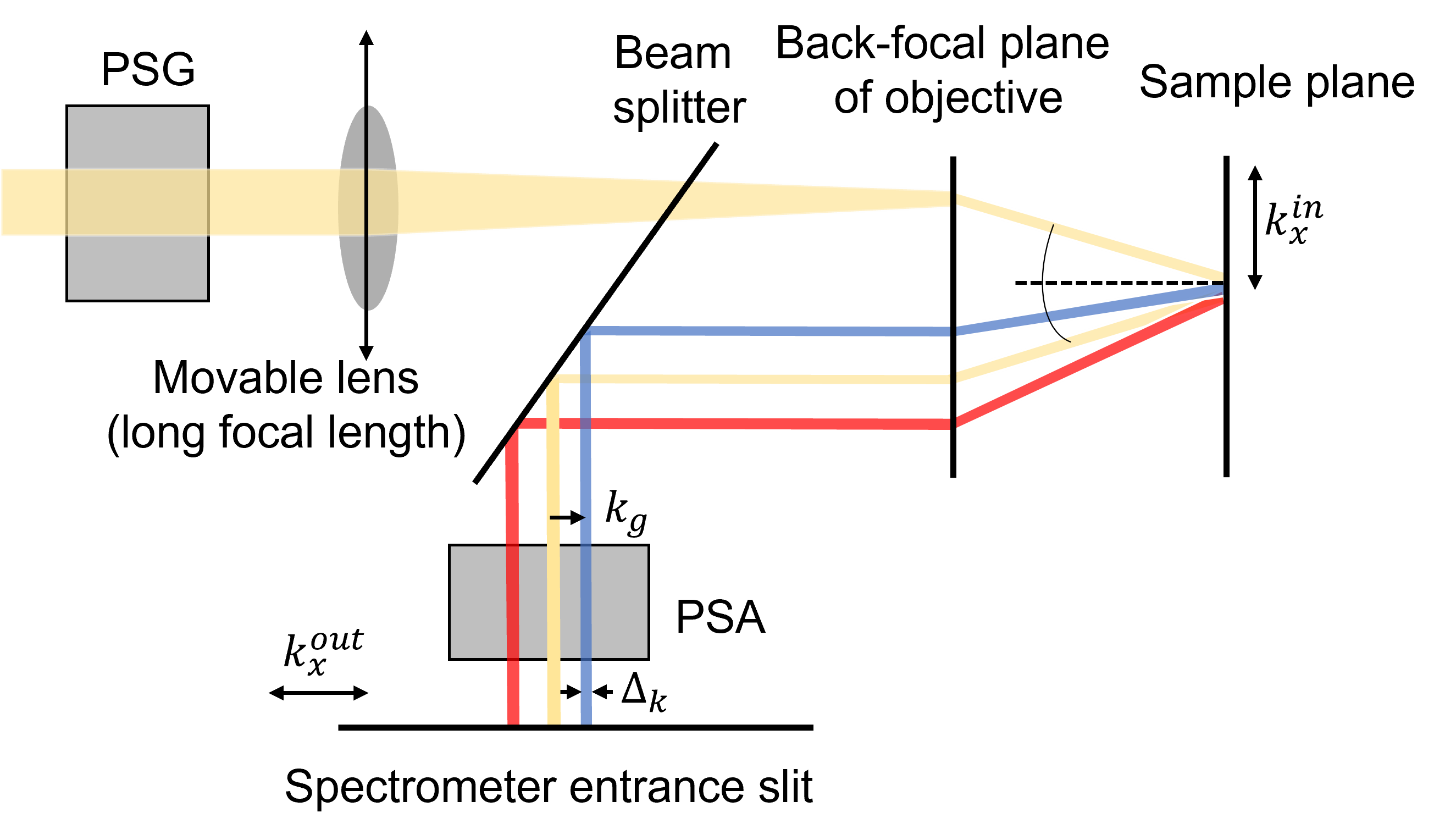}
\caption{Schematics of the experimental setup developed for acquiring Mueller matrices that connect the input to output Stokes vectors associated with the polarization states incident on the metasurface (`sample' in the schematics) and reflected from it. Stokes vectors ${\bf S}=(I_0,I_V-I_H,I_{45}-I_{-45}, I_+-I_-)$ gather the total intensity of the beam with three balanced intensity measurements performed with vertical ($V$), horizontal ($H$), $+45^\circ$-, $-45^\circ$-tilted linear polarizations and $+$ and $-$ circular polarizations. Input Stokes vectors are prepared at the PSG stage through which the incident beam is transmitted. Polarization states of the reflected light are analyzed as output Stokes vectors through the PSA. By focusing the incident beam on the back-focal plane of the illumination objective, the movable lens allows for measuring Mueller coefficients in the $(\omega,k_x^{in})$ parameter space, as detailed in the text and Fig. \ref{fig:exp_meas} below. In the reflection Fourier space $k_x^{out}$, the polarization states of the scattering processes (here separated from the specular reflection by one geometric momentum $\pm k_g$) are spectrally analyzed.}
\label{fig:exp_setup}
\end{figure}

Mueller polarimetry is the appropriate tool for measuring artifact-free polarization responses of general optical systems, such as plasmonic metasurfaces. The matrix connects input to output Stokes vectors (representing the incident's polarization states and scattered light fields). As such, the Mueller matrix is a $4\times 4$ matrix whose coefficients can be determined following well-known methodologies we implemented in previous work \cite{Thomas2018,Lorchat2018}. Here, we extend these methodologies to measure angularly-resolved Mueller coefficients (i.e., measuring the Mueller matrix in the Fourier space) as a function of both the photon energy $\omega$ and the incoming momentum $k_x^{in}$ (same frame as defined in the main text, see Fig. 1 there). 

The experimental setup is schematized in Fig. \ref{fig:exp_setup} where the incident polarization states of light are \textit{prepared} using a combination of one polarizer and a quarter-wave plate (gathered within the Polarization State Generator stage -PSG- in the Figure), and the output polarization states are \textit{analyzed} with an identical combination of polarization optics (gathered within the Polarization State Analysis stage -PSA- in the Figure). As explained in detail in \cite{LiPhD}, our measurements collect the far-field intensities for all positions of the PSG and PSA necessary to fill in the 16 coefficients of the Mueller matrix.

\begin{figure}[ht]
\includegraphics[width=0.8\columnwidth]{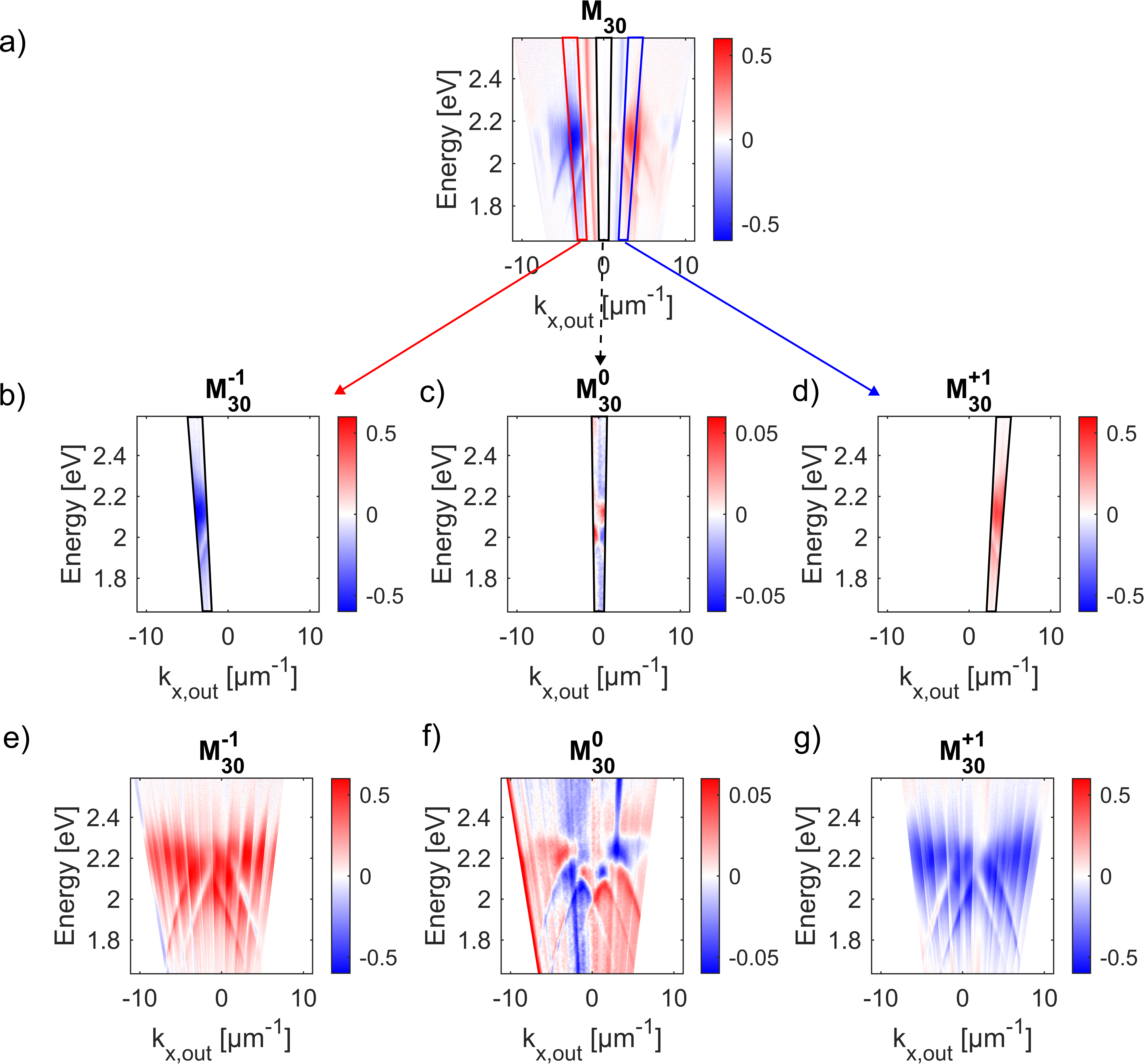}
\caption{To measure the different $M^{n}_{30}(\omega, k_x^{out})$ coefficients, we start with one lateral position of the movable lens shown in Fig.~\ref{fig:exp_setup} above. This leads to illuminating the metasurface with a given in-plane momentum $k_x^{in}$ within an angular resolution $\Delta_{k}$ schematized by the rectangular boxes in panel (a). Under such a fixed $k_x^{in}$, the different diffraction orders can be easily identified in the $(\omega, k_x^{out})$ Fourier space shown in panel (a). As described in the lower panels (b), (c), and (d), each diffraction order $k_x^{in} \pm k_{g}$ including specular reflection, is spectrally analyzed by scanning the available $k_x^{in}$ angular spectrum in steps of $\Delta_{k}$. The spectral data are then stitched one to the other to form the Mueller $M_{30}$ coefficients in the full $(\omega, k_x^{out})$ space. We need to correct this raw data for the polarimetric contribution of the other optical element using Eq.~\ref{Eq:MMFullycorrected}. The resulting spectra for the $-1$, $0$ and $+1$ orders are indicated in panels (e), (f), and (g), respectively.}
\label{fig:exp_meas}
\end{figure}

In the main text, we focus on the $M_{30}$ Mueller coefficient as it quantifies the ratio between the $\pm$ circular polarization contrast $I^+_{out}-I^-_{out}$ in the far field and the total intensity of the light reflected from the metasurface. Fig.~\ref{fig:exp_meas} displays the experimental construction for the $M^{n_1+n_2}_{30}$ elements that are discussed in the main text for the three processes that fulfill $k_x^{out}= k_x^{in}+(n_1+n_2)k_g$ -see main text. The $n=n_1+n_2=0$ case (specular reflection case) yields the three $(n_1,n_2)$ pairs of branches $(0,0)$, $(-1,1)$, and $(1,-1)$ that appear on $M^0$. For $M^{\pm 1}$, two combinations only are measured with $(\pm 1,0)$ and $(0,\pm 1)$ and only one process $(\pm 1,\pm 1)$ observed for $M^{\pm 2}$, considering that $M^{+n}_{30}(k)=-M^{-n}_{30}(-k)$.

Laterally shifting the excitation beam incident on the illuminating objective, the incoming momentum is selected using a movable lens. This selection of a given in-plane momentum is set within an angular resolution $k_x^{in}±\pm\Delta_k / 2$ fixed by the numerical aperture of the shifting lens together with its focal length. For a given $k_x^{in}$, we measure the whole $k_x^{out}$ spectrum available by the numerical aperture of the objective used. We then measure the Mueller matrix in the full $(\omega, k_x^{out})$ space by adjusting the PSG and PSA states accordingly. We scan $k_x^{in}$ through the same available angular spectrum in 13 steps so that, between each measurement, the beam is approximately shifted by an equivalent $\Delta_k$ in the back-focal plane. In panel (a) of Fig. \ref{fig:exp_meas}, we display the $M_{30}$ coefficients in the entire $(\omega,k_x^{out})$ space for one specific $k_x^{in}$. In panels (b), (c), and (d), we thereby obtain, for successive $k_x^{in}$ values, spectrally analyzed Mueller $M_{30}$ coefficients recorded over ``stripes'' of angular spread $\Delta_k$ at $k_x^{out}=k_x^{in}+nk_g$ for $n=0, \pm1$. 

The obtained Mueller matrix (MM) must be corrected for different polarimetric contributions in our optical setup, as shown in Fig.~\ref{fig:exp_setup}. The first step is to take into account beam splitter corrections. We denote the MM of the beam splitter reflection as $\textbf{M}_{BSR}$ and its transmission  MM as $\textbf{M}_{BST}$ such as the experimental MM, denoted $\textbf{M}_{EXP}$, can be written as a function of the MM of the sample, denoted  $\textbf{M}_{S}$. The final step to correct our signal is to take into account the change of referential due to the mirror reflection. To this end, we use the MM shown in Eq.~\ref{Eq:MMMirror}, which represent a perfect mirror,

\begin{align}     
\textbf{M}_{R}=\begin{pmatrix}
1 & 0 & 0 & 0\\ 0 & 1 & 0 & 0\\ 0 & 0 & -1 & 0\\0 & 0 & 0 & -1
\end{pmatrix}.
\label{Eq:MMMirror}
\end{align}

The experimental MM matrix can then be written in serial decomposition, such as:
\begin{align}   
\textbf{M}_{EXP}= \textbf{M}_{BST}\textbf{M}_{R}\textbf{M}_{S}\textbf{M}_{BSR}.
\label{Eq:MMFullycorrected}
\end{align}

Using Eq.~\ref{Eq:MMFullycorrected}, we can correct the experimental MM by multiplying the experimental MM by the inverse MM of each optical element in the correct order. The resulting $M^{n}_{30}(\omega, k_{x}^{out})$ coefficients are then shown in the panel (e), (f) and (g) of Fig.~\ref{fig:exp_meas} respectively. We then associate for each order at a given $k_{x}^{out}$, a given $k_{x}^{in}$ to obtain the spectra shown in the main text, as $k_{x}^{out}=k_{x}^{in}+nk_{g}$.

